\documentclass[linenumbers]{aastex631}

\usepackage{amsmath}
\usepackage{graphicx}
\usepackage{subfigure}

\def\GHz{\mathrm{GHz}}

\begin{document}

\title{Inferring the Intrinsic Energy Function of FRB 20220912A}

\author[0000-0002-2552-7277]{Xiaohui Liu}

\affiliation{National Astronomical Observatories, Chinese Academy of Sciences, Beijing 100101, China}
\affiliation{University of Chinese Academy of Sciences, Beijing 100049, China}

\author[0000-0001-9036-8543]{Wei-Yang Wang}
\affiliation{University of Chinese Academy of Sciences, Beijing 100049, China}
\email{wywang@ucas.ac.cn}

\author[0000-0002-1056-5895]{Weicong Jing}
\affiliation{National Astronomical Observatories, Chinese Academy of Sciences, Beijing 100101, China}
\affiliation{University of Chinese Academy of Sciences, Beijing 100049, China}

\author[0000-0001-6475-8863]{Xuelei Chen}
\affiliation{National Astronomical Observatories, Chinese Academy of Sciences, Beijing 100101, China}
\affiliation{University of Chinese Academy of Sciences, Beijing 100049, China}
\affiliation{Key Laboratory of Radio Astronomy and Technology, CAS, Beijing 100101, China}
%\affiliation{State Key Laboratory of Radio Astronomy and Technology, Beijing 100101, People's Republic of China}
\email{xuelei@cosmology.bao.ac.cn}

\author[0000-0002-9274-3092]{Jinlin Han}
\affiliation{National Astronomical Observatories, Chinese Academy of Sciences, Beijing 100101, China}
\affiliation{University of Chinese Academy of Sciences, Beijing 100049, China}


\begin{abstract}
The statistical analysis of fast radio burst (FRB) samples from repeaters may suffer from a band-limited selection effect, which can bias the observed distribution. We investigated the impact of this selection bias on the energy function through simulations and then applied our analysis to the particular case of FRB 20220912A.
Our simulations show that, in the sample of bursts observed by the Five hundred meter Aperture Spherical Telescope (FAST), assuming a unimodal intrinsic energy distribution, the band selection effect alone is insufficient to produce a bimodal energy distribution; only the bimodal central frequency distribution can achieve this.
The bursts' energy of FRB 20220912A that primarily fell within the observing band showed no significant correlation with the central frequency.
In contrast, bursts with higher central frequency tend to exhibit narrower bandwidth and longer duration.
The distribution of the intrinsic energy can be modeled as a log-normal distribution with a characteristic energy of $8.13 \times 10^{37}$ erg, and a power-law function with the index of $1.011 \pm 0.028$. In contrast to the initial energy function reported by \cite{2023ApJ...955..142Z}, the low-energy peak vanishes, and the high-energy decline becomes steeper, which implies the low-energy peak is an observational effect.
The bimodality of the energy distribution seems to originate from the intrinsic radiation mechanism.
\end{abstract}



\section{Introduction} \label{sec:intro}
FRBs are intense radio signals that typically last only for milliseconds and generally come from the distant universe beyond our Galaxy \citep{2014ApJ...790..101S, 2017ApJ...834L...7T, 2017ApJ...834L...8M, 2017Natur.541...58C, 2020Natur.581..391M}.
The detection of an FRB-like burst from a Galactic magnetar has suggested a potential connection between FRBs and magnetars \citep{2020Natur.587...59B, 2020Natur.587...54C}.
The large samples collected from different sources suggest that there may be two distinct populations of FRBs \citep{2021ApJS..257...59C}, namely the repeaters and the one-off bursts, and the repeaters seem to be characterized by longer durations and narrower bandwidths \citep{2021ApJ...923....1P}, though the existence of two distinct types of FRBs is still a subject of ongoing debate \citep{2018ApJ...854L..12P, 2018NatAs...2..839C, 2023PASA...40...57J, 2024ApJ...975...75L, 2025ApJ...993...37B, 2025ApJ...989..144L, 2025MNRAS.540.3709U}.

Various temporal-frequency structures, such as downward or upward drifting, or more complex structures, were found in a sample of more than a few hundred burst events \citep{2022RAA....22l4001Z}.
The sample indicates a large energy budget \citep{2024arXiv240507152W}, and the energy distribution of the repeaters provides crucial insights into their origin.
The energy function of the bright sample from individual sources \citep{2017ApJ...850...76L, 2019ApJ...882..108W, 2019ApJ...877L..19G, 2021MNRAS.500..448C, 2021ApJ...922..115A, 2020Natur.582..351C} can be modeled by the Schechter function \citep{1976ApJ...203..297S}, which is characterized by a power law function with an exponential cutoff.
However, the inclusion of low-energy outbursts requires a two-component energy distribution to fit the data \citep{2021Natur.598..267L, 2022RAA....22l4002Z, 2023ApJ...955..142Z, 2022Natur.611E..12X, 2022Natur.611E..10N, 2023MNRAS.519..666J, Zhang:2025qzn}, and the physics behind it is still unknown.

FRB 20220912A, discovered by the CHIME/FRB collaboration \citep{2022ATel15679....1M}, is a continuously active source with a clean and stable surrounding environment \citep{2024ApJ...974..296F, 2023ApJ...955..142Z}. The Deep Synoptic Array (DSA-110) localized this source in a host galaxy with a redshift of $0.077$ \citep{2023ApJ...949L...3R}, and its location was later refined to milliarcsecond precision by the European VLBI Network \citep{2024MNRAS.529.1814H}.
The Allen Telescope Array utilized the advantageous 2.7 GHz observing band to detect bursts, and the correlations between the morphological parameters of these bursts have been studied \citep{2024MNRAS.52710425S}. \cite{2025MNRAS.tmp.1851O} also conducted about 1500 h of observations to obtain a sample of bright bursts and probe the maximum energy of FRB.
Within a period of $8.67$ hours, 1076 bursts were detected by FAST and the highest event rate can be up to $390~ \mathrm{hr^{-1}}$ above the fluence threshold 0.0148 Jy ms \citep{2023ApJ...955..142Z}.
The detected burst spectra are found to fall within an extremely narrow bandwidth, with the log-mean value of $\Delta \nu / \nu \sim$ 0.17. While propagation effects, such as absorption, scintillation, and plasma lensing in the source environment may lead to narrow spectra, the probability of these phenomena being the cause is very low \citep{2023ApJ...956...67Y, 2024ApJ...974..160K}.
The narrow spectra could also be a consequence of intrinsic radiation mechanisms, such as quasi-periodic distribution in the bulk of bunches or the negative absorption in the maser mechanism \citep{2023ApJ...956...67Y, 2024A&A...685A..87W}.

The energy functions of repeaters exhibit distinctive features that provide crucial insights into their emission mechanism.
For example, the similarity between the energy functions of FRBs and giant pulses of pulsars may imply a common origin \citep{2019MNRAS.490L..12B, 2020MNRAS.491.1498C}. However, a direct analogy is challenged by the substantial gap in their absolute luminosities, complicating a simple scaling relation. Furthermore, the scenario is complicated by the observation of a bimodal energy function in some active repeating FRBs \citep{2021Natur.598..267L, 2022RAA....22l4002Z, 2023ApJ...955..142Z, 2022Natur.611E..12X, 2022Natur.611E..10N, 2023MNRAS.519..666J, Zhang:2025qzn}.
The precise estimation of the burst energy is dependent on the spectral shape.
For broadband emission (e.g., pulsars and Lorimer burst, \cite{2007Sci...318..777L}), where most of its energy is beyond the bandwidth of the telescope, the product of the central frequency $\nu_{\mathrm{c}}$ and the specific fluence is more appropriate \citep{2023RvMP...95c5005Z}.
Conversely, for narrowband emission, the energy is typically calculated using the observing bandwidth $\mathrm{BW}_{\mathrm{fil}}$ rather than $\nu_{\mathrm{c}}$.
This method yields an accurate estimation if the entire spectrum falls within the observing bandwidth $\mathrm{BW}_{\mathrm{fil}}$, but would underestimate the true energy if the radiating bandwidth is only partially covered.
Critically, the $\nu_{\mathrm{c}}$ distributions are often clustered near the band edges of the telescope \citep{2022RAA....22l4001Z,2023ApJ...955..142Z}.
Such a tendency would amplify the bias in energy estimation and distort the inferred energy function by systematically underestimating a significant fraction of bursts.
Additionally, if burst energy correlates with central frequency, the observed sample may further deviate from the true energy distribution.
Both scenarios must be carefully examined before drawing conclusions based on the observed energy function.

This article investigates whether the observed bimodal energy distribution of FRB 20220912A is of physical origin, or caused by the incomplete frequency coverage in observation. We will study two possible scenarios where the pseudo-bimodal energy distribution could be generated: the band-limited selection effect, and the population features, especially the correlation between the energy and the central frequency.
To achieve our objectives, we conduct a detailed simulation to examine the impact of band selection effects on the energy distribution and use the observational data of FRB 20220912A as a case study to analyze the population features.
Finally, we study how to mitigate the above effects in the reconstruction of the intrinsic energy distribution.

This paper is organized as follows.
In Section \ref{sec:met} we describe the method and the data used in this paper.
In Section \ref{sec:se} we investigate the impact of the band-limited selection effect.
In Section \ref{sec:pf} we investigate the population features of FRB 20220912A.
In Section \ref{sec:intE} we reconstruct the intrinsic energy distribution.
We discuss some implications in Section \ref{sec:dis} and conclude our results in Section \ref{sec:con}.

\section{Method} \label{sec:met}
\subsection{Burst Model}
The dynamic spectra of FRB repeaters show the intensity across the frequency and time.
Unlike the non-repeaters which usually have broadband spectra, the repeaters typically have a narrower bandwidth \citep{2021AAS...23723603P}, whose spectra can be well modeled as a Gaussian function \citep{2017ApJ...850...76L}.
This empirical model is validated experimentally for many sources \citep{2021ApJ...922..115A, 2021Natur.598..267L, 2022RAA....22l4001Z, 2022RAA....22l4002Z, 2022Natur.611E..12X, 2022Natur.611E..10N, 2023ApJ...955..142Z, 2023MNRAS.519..666J}.
We assume that the time-integrated spectra could be modeled as a Gaussian distribution,
\begin{equation}
    F(f) = F_{\mathrm{int}} \mathrm{G} (f | \nu_{0}, \sigma),
\end{equation}
where $f$ is the frequency, $F_{\mathrm{int}}$ is the intrinsic fluence of this burst, $\nu_{0}$ and $\sigma$ represent the central frequency and width of the Gaussian spectra, respectively. The Gaussian function is defined as $\mathrm{G} (f | \mu, \sigma) = \frac{1}{2 \pi \sigma} e^{-\frac{\left( f-\mu \right)^2}{2\sigma^2}}$. It is also convenient to use the full-width-at-half-maximum (FWHM) bandwidth, $\mathrm{BW}_{50} = 2\sqrt{2 \ln{2}}~ \sigma$.
Due to the limited observing band of a telescope, the observed fluence typically covers only a fraction of the intrinsic fluence, 
\begin{equation}\label{eq:obsint}
    F_{\mathrm{obs}} = F_{\mathrm{int}} \int_{\nu_{l}}^{\nu_{u}} \mathrm{G} (f | \nu_{0}, \sigma) df,
\end{equation}
where $\nu_{l}$ and $\nu_{u}$ are the lower and upper limits of the observing band, respectively, and
$\mathrm{BW}_{\mathrm{fil}} = \nu_{u} - \nu_{l}$ is the width of the observing band \footnote{This definition of fluence differs from that in \cite{2023ApJ...955..142Z} by a factor of $\mathrm{BW}_{\mathrm{fil}}$.}.
The integration in Equation (\ref{eq:obsint}) should exclude the contaminated frequencies due to the Radio Frequency Interference (RFI). Here we assume the contaminated bands are narrow and can be neglected in the case of FAST observation of FRB 20220912A.
We relate the fluence to the isotropic-equivalent energy by
\begin{equation} \label{E2S}
    E_{\mathrm{int}} = 10^{39}\,\mathrm{erg} \left(\frac{4 \pi}{1+z} \right)\left( \frac{D_L}{10^{28} \mathrm{~cm}} \right)^2 \left(\frac{F_{\mathrm{int}}}{\mathrm{Jy} \cdot \mathrm{ms} \cdot \mathrm{GHz}}\right),
\end{equation}
where $z$ and $D_L$ represent the redshift and the luminosity distance of the source, respectively.
We adopted the redshift $z = 0.0771$ and the luminosity distance $D_L = 360.86\,\mathrm{Mpc}$ for FRB 20220912A \citep{2023ApJ...949L...3R} as the fiducial values in the simulation.
The intrinsic energy $E_{\mathrm{int}}$ defined this way represents the energy after the correction of the limited bandwidth, without consideration of propagation effects and other factors.

\subsection{FRB Population and Energy Functions}
From the de-dispersed dynamic spectra, the arrival time, the fluence within $\mathrm{BW}_{\mathrm{fil}}$, and the parameters characterizing the dynamic spectra morphology are estimated. The joint probability density function (PDF) is given by $p(E_{\mathrm{int}},\boldsymbol{\theta})$, where $\boldsymbol{\theta}$ is the vector of all morphological parameters. For a sample of bursts originating from a single source, the bursts are likely dominated by a certain physical process, which is reflected by its intrinsic properties, such as the distribution of morphological parameters or correlations among such parameters.

In the pulse search, a burst is identified when the ratio of the observed average flux density to the system noise exceeds a chosen threshold $\mathrm{SNR}_{\mathrm{th}}$.
The signal-to-noise ratio (SNR) of a burst is given by
\begin{equation} \label{SNR}
    \mathrm{SNR} = \frac{F_{\mathrm{obs}}}{\mathrm{BW}_{\mathrm{fil}} \text{ } \mathrm{W}_{\mathrm{eq}}} \times \frac{G \sqrt{n_{\mathrm{p}} \text{ } \mathrm{BW}_{\mathrm{fil}} \text{ } \mathrm{W}_{\mathrm{eq}}}}{T_{\mathrm{sys}}},
\end{equation}
where $T_{\mathrm{sys}}$ and $G$ are the telescope's system temperature and gain, which are 20 K and 16.1 K/Jy, respectively, for FAST,
$n_{\mathrm{p}} = 2$ is the number of polarization channels, and $\mathrm{BW}_{\mathrm{fil}} = 0.5$ GHz for the FAST L-band receiver.
$\mathrm{W}_{\mathrm{eq}}$ is the equivalent duration of the FRB signal, which is usually defined by dividing the fluence by the peak flux.
When we take into account the observational effect, the PDF can be written as
\begin{equation} \label{eq:pEdot}
    p'(E_{\mathrm{int}},\boldsymbol{\theta}) = \frac{1}{C} p(E_{\mathrm{int}},\boldsymbol{\theta}) S(E_{\mathrm{int}},\boldsymbol{\theta}),
\end{equation}
where $S(E_\mathrm{int},\boldsymbol{\theta})$ is the selection function, which describes whether there is a sufficiently high signal-to-noise ratio within the observing band that the telescope can detect.
To maintain consistency with previous works, we have chosen the SNR threshold $\mathrm{SNR}_{\mathrm{th}}$ to be 7.
Thus, for bursts exceeding the threshold $ \mathrm{SNR} \ge 7$, we have $S(E_\mathrm{int}, \boldsymbol{\theta}) = 1 $ and $ S(E_\mathrm{int}, \boldsymbol{\theta} ) = 0 $ otherwise. $C$ is the normalization factor. Real data are often affected by extensive RFI, and the residual RFI after masking can contribute to a non-Gaussian noise. Additionally, the discrete trail of dispersion measure (DM) in the pulse search stage may miss some candidates. However, for the observations of FAST, the off-line pulse search usually uses a very dense trail, and the candidates are cross-checked by different members, which can effectively reduce the risk of raising the real SNR threshold. Furthermore, the observed sample contains very few bursts with an SNR precisely in the range of 7 to 8 \citep{2023ApJ...955..142Z, 2022Natur.611E..12X}. Thus, adopting a threshold of 7 is statistically robust.
We clarify that while the machine-learning algorithms can enhance burst detection at lower SNRs \citep{2025arXiv251007002W}, such methods primarily recover bursts that are fainter or narrower in bandwidth than those detectable by traditional SNR threshold and do not significantly affect the completeness of our sample above this limit.

Reliable reconstruction of intrinsic parameters from the dynamic spectra becomes challenging when only a portion of the spectra is observed.
Relying on the observed energy to infer the energy function may lead to an additional bias.
The joint PDF of the observed energy $E_{\mathrm{obs}}$ and the morphological parameters $\boldsymbol{\theta}$ can be written as
\begin{equation} \label{eq:pEobs}
    p_{\mathrm{obs}}(E_{\mathrm{obs}},\boldsymbol{\theta}) = p'(E_{\mathrm{obs}},\boldsymbol{\theta}) J(\boldsymbol{\theta}),
\end{equation}
where $J = (\int_{\nu_{b}}^{\nu_{u}} \mathrm{G} (f | \nu_{0}, \sigma) df)^{-1}$ is the determinant of the Jacobian matrix for the change of variable $E_{\mathrm{int}} \rightarrow E_{\mathrm{obs}}$. Note that in marginalizing the morphological parameters in $p(E_{\mathrm{int}}, \boldsymbol{\theta})$, Equation (\ref{eq:pEdot}) and Equation (\ref{eq:pEobs}), the 
true energy function $p(E_{\mathrm{int}})$, the PDF 
$p'(E_{\mathrm{int}})$ and the observed $p_{\mathrm{obs}}(E_{\mathrm{obs}})$ are each different, due to the selection function $S$ and the Jacobian $J$.

\subsection{Simulation Procedure}
We carry out a Monte Carlo simulation to evaluate these energy functions.
For each simulation, we generate more than $10^8$ mock bursts with their $E_{\mathrm{int}}$, $\mathrm{W}_{\mathrm{eq}}$, $\nu_0$ and $\mathrm{BW}_{50}$ drawn from the predefined PDFs. 
Then, the observed energy and SNR can be derived from Equation (\ref{eq:obsint}), Equation (\ref{E2S}), and Equation (\ref{SNR}).
After removing the bursts with SNR $< 7$ in the sample, $p'(E_{\mathrm{int}})$ and $p_{\mathrm{obs}}(E_{\mathrm{obs}})$ are obtained.
Note that we have simplified the simulation process, neglecting additional biases from pulse search, de-dispersion, and fitting procedures. 

Throughout this paper, we assume that  $\mathrm{BW}_{50}$/GHz follows a log-normal distribution $\mathcal{LN} (0.251, 0.349)$, and the $\mathrm{W}_{\mathrm{eq}}$/ms distribution follows a log-normal distribution $\mathcal{LN} (4.67, 0.64)$, matching the case of FRB 20220912A \citep{2023ApJ...955..142Z}.
The lognormal function is expressed as
\begin{equation}
\mathcal{LN} (x | \mu, \sigma) = \frac{1}{\sqrt{2 \pi} x \sigma} e^{- \frac{(\log(x) - \mu)^2}{2 \sigma^2}} 
\end{equation}
As for the intrinsic energy distribution, we considered two typical energy models: log-uniform distribution $\log_{10}(E_{\mathrm{int}}/\mathrm{erg}) \sim \mathcal{U}(10^{35},10^{40})$, and log-normal distribution $E_{\mathrm{int}}/\mathrm{erg} \sim \mathcal{LN} (\log(10^{38}), 1.2)$.
It is convenient to use this log-uniform toy model to study the impact of the band selection effect.
In this scenario, the completeness, defined as the fraction of detected bursts to the simulated bursts at a specific central frequency, is analogous to $p'(E_{\mathrm{int}})$.

So far, there is no reliable model available for the burst rates as a function of frequency, so we consider some simple $\nu_{0}$ (GHz) distributions.
The unimodal distributions we considered include the Gaussian distribution, the Lorentz distribution, and the Exponential distribution,
\begin{equation}
\begin{aligned}
& \text{Gaussian}: p(\nu_{0}) = \mathrm{G}(\nu_{0} | \nu_{1}, \sigma), \\
& \text{Lorentz}: p(\nu_{0}) = \frac{1}{\pi} \left[ \frac{\gamma}{(\nu_{0} - \nu_{1})^2 + \gamma^2} \right], \\
& \text{Exponential}: p(\nu_{0}) = \lambda \exp{ \left( -\lambda \left( \nu_{0} - \nu_{1} \right) \right)},
\end{aligned}
\end{equation}
where for economy of symbol we used $\nu_{1}$ to represent the typical frequency in the different distributions, though strictly speaking each of these is a different (unrelated) parameter.
The parameters $\sigma$, $\gamma$, and $\lambda$ are the shape parameters that can control the width of the distributions.

The bimodal central frequency distribution we consider is a sum of two Gaussian functions, with one given a central frequency distributed inside the observing band, and one outside, so that 
\begin{equation}
    p(\nu_0) = \alpha \mathrm{G} (\nu_0 | \nu_{1}, \sigma_{1}) + (1 - \alpha) \mathrm{G} (\nu_0 | \nu_{2}, \sigma_{2}),
\end{equation}
where $\alpha$ is the fraction of the first Gaussian function, and \{$\nu_{1}, \sigma_{1}$, $\nu_{2}, \sigma_{2}$\} are the parameters related to the shape of $\nu_0$ distribution.
In fact, we have also experimented with a variety of parameter combinations, but all yield similar results.
Therefore, we only show one typical case with $\sigma_1 = \sigma_2 = 0.1$ GHz and $\nu_2 = 1.25$ GHz.

\subsection{Evaluating the Population Features}
The data used in this paper includes the frequency-averaged fluence $F_{\mathrm{obs}} / \mathrm{BW}_{\mathrm{fil}}$, the central frequency $\nu_0$, the bandwidth $\mathrm{BW}_{50}$, and the duration $\mathrm{W}_{50}$ of FRB 20220912A \citep{2023ApJ...955..142Z}.
The data points with the error of bandwidth larger than half the width of the observing band $\mathrm{BW}_{\mathrm{fil}}/2 = 0.25 \mathrm{GHz}$ are removed due to their large uncertainty.
In principle, the distributions of these parameters are coupled with each other, but within a narrow observing band, the correlation between $\nu_0$ and the other parameters can be treated as linear.
These features can be used to examine potential biases in the sample.

Bursts with $\nu_{0}$ near the center of the observing band 
do not suffer significantly from the band selection effect, while those near the edge are much affected.
Considering a sub-sample with $\nu_{0}$ ranging from $1.25 - \Delta f ~ \GHz$ to $1.25 + \Delta f ~ \GHz$, we gradually increased $\Delta f$ and then conducted linear fitting on each sub-sample. Given that these quantities usually span many orders, it is more practical to perform a linear fit using their base-10 logarithms, and the slope $k$ of them against $\nu_0$ is defined as $\log_{10}(y) \propto k_{\log_{10}(y)} \nu_0$, where $y$ can be $E_{\mathrm{int}}$, $\mathrm{BW}_{50}$ and $\mathrm{W}_{\mathrm{eq}}$.
To account for the error of variables, the bootstrap method is used.
For each fit, we first draw each value of the parameters from a normal distribution, then we perform a linear fit and repeat the entire process $10^4$ times. The best-fit slope $k_i$ obtained from each process is collected to be the posterior distribution of the true slope $k$, and we report the mean and the standard deviation of $k$.
During the sampling process, the non-physical realizations are removed by requiring that the values sampled from the distribution in each process must be positive.

\section{Band Selection Effects} \label{sec:se}
\subsection{The Influence of Band-limited Selection Effects}
\begin{figure}
    \centering
    \includegraphics[width=0.9\linewidth]{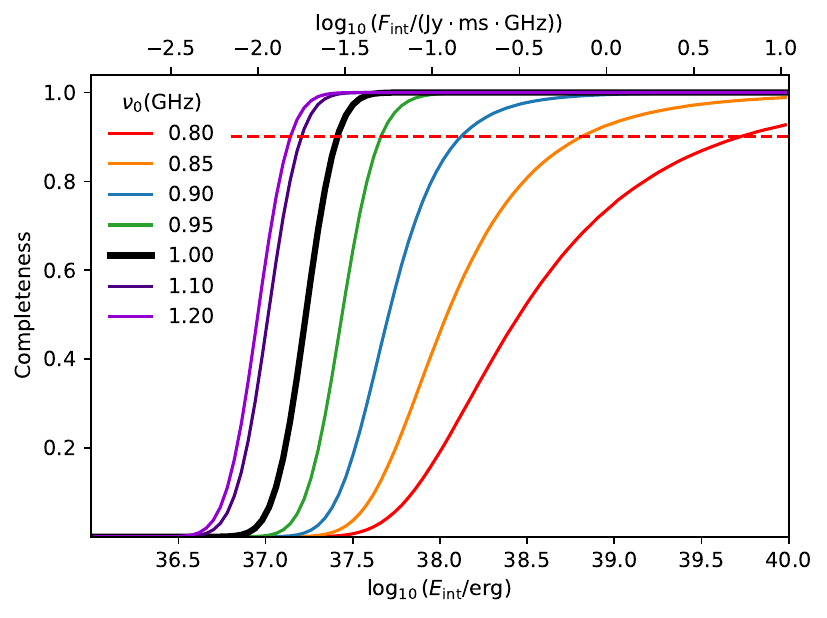}
    \caption{Completeness function for different values of $\nu_{0}$, which shows the fraction of bursts that can be detected in the simulation. The black bold line is the scenario where $\nu_{0}$ aligns with the band edge ($1 \mathrm{GHz}$), resulting in only half of the spectrum being received. The red dotted line represents the 90\% completeness threshold.}
    \label{fig: CCF}
\end{figure}

The completeness as a function of the burst energy is shown for different $\nu_0$ scenarios in Figure \ref{fig: CCF}.
When $\nu_0$ falls within the observing band, the completeness function has a sharp cutoff at the low energy threshold.
Above the detection limit, the completeness is nearly unity, indicating that bursts above the threshold energy are almost certain to be detected.
This tail only shifts slightly as long as $\nu_0$ is within the observing band.

For Gaussian burst spectra, only a subset of bursts can be detected when the central frequency lies outside the observational band, resulting in a sharp decline in detection completeness.
Naturally, higher-energy bursts exhibit greater completeness than their lower-energy counterparts.
In practice, there is no well-defined boundary between complete and incomplete sections.
Current methodologies typically address this by simulating a characteristic energy threshold, $E_{90}$, defined as the energy above which the detection completeness reaches 90\%.
The 90\% completeness threshold $E_{90}$ is used as an indicator to show that the energy function above $E_{90}$ is complete in previous works \citep{2021Natur.598..267L, 2022RAA....22l4002Z}.
Our simulations further reveal that $E_{90}$ is frequency-dependent.
Consequently, if the central frequency distribution of an observed sample is uncertain, the derived $E_{90}$ may be significantly inaccurate.

\begin{figure}
    \centering
    \includegraphics[width=0.9\linewidth]{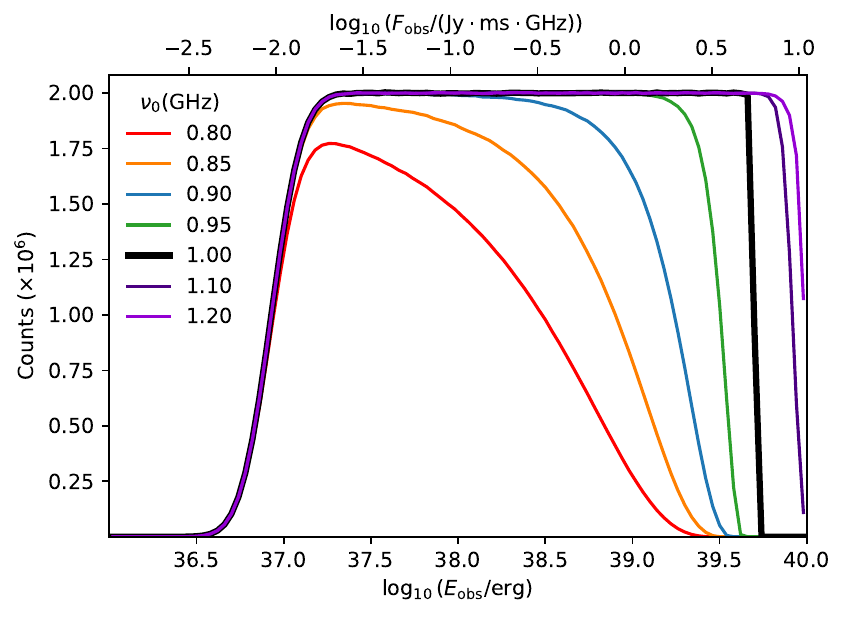}
    \caption{The observed energy distributions $p_{\mathrm{obs}}(E_{\mathrm{obs}})$ for different values of $\nu_{0}$. The black bold line is the scenario where $\nu_{0}$ aligns with the band edge ($1 \mathrm{GHz}$), resulting in only half of the spectrum being received.}
    \label{fig: EobsCF}
\end{figure}

The observed energy distributions of different $\nu_0$ scenarios are shown in Figure \ref{fig: EobsCF}.
The observed energy function $p_{\mathrm{obs}}(E_{\mathrm{obs}})$ shows similar behavior to the completeness function.
When the spectra fall primarily within the observing band, the determinant term $J(\boldsymbol{\theta})$ is trivial, and the distribution of the observed energy remains uniform within the majority of energy ranges we are considering.
Note that, even if the PDF of the observed energy is uniform, some deviations may occur due to the non-matched bandwidth, resulting in discrepancies between the observed energy and the intrinsic energy. 
This is also why deviations from a uniform distribution can be observed at the high-energy end of the curve.
After the $\nu_0$ moves out of the observing band, $p_{\mathrm{obs}}(E_{\mathrm{obs}})$ is no longer uniform, but instead exhibits a peak near the detection threshold.
In this case, the bursts with high energy may be detectable, but the apparent energy is significantly reduced. 

In conclusion, the band selection effects induced by the selection function $S$  bias the distribution of the sample.
When $\nu_0$ falls within the observing band, the band selection effects suppress the completeness function near the detection limit, and the observed energy can keep the log-uniform shape except for the high-energy tail.
In contrast, if $\nu_0$ is outside the band,  the completeness decreases drastically, and the observed energy distribution exhibits a peak.

\subsection{Unimodal Central Frequency Distributions}
\begin{figure*}
\centering
\subfigure[Gaussian]{\includegraphics[width=2in]{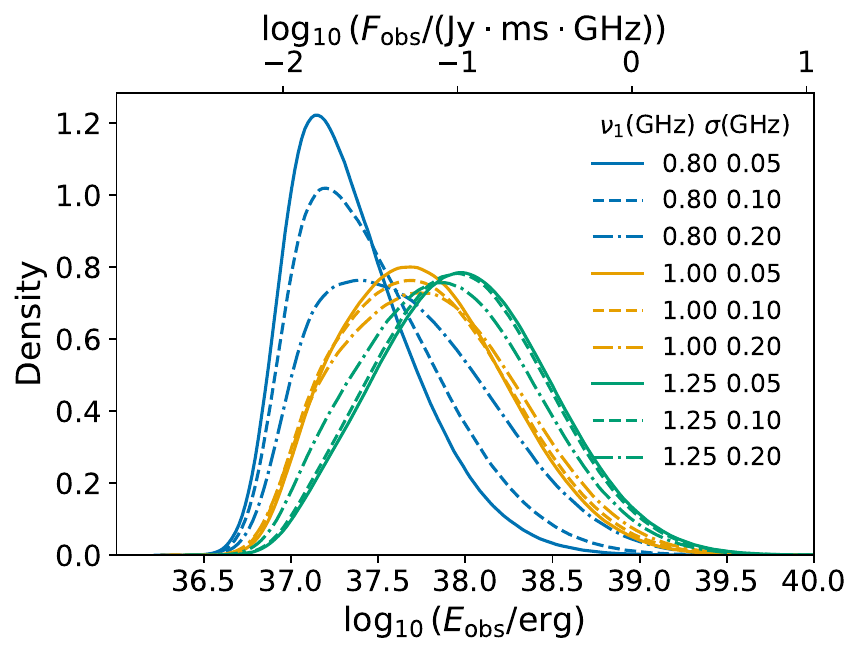}}
\subfigure[Lorentz]{\includegraphics[width=2in]{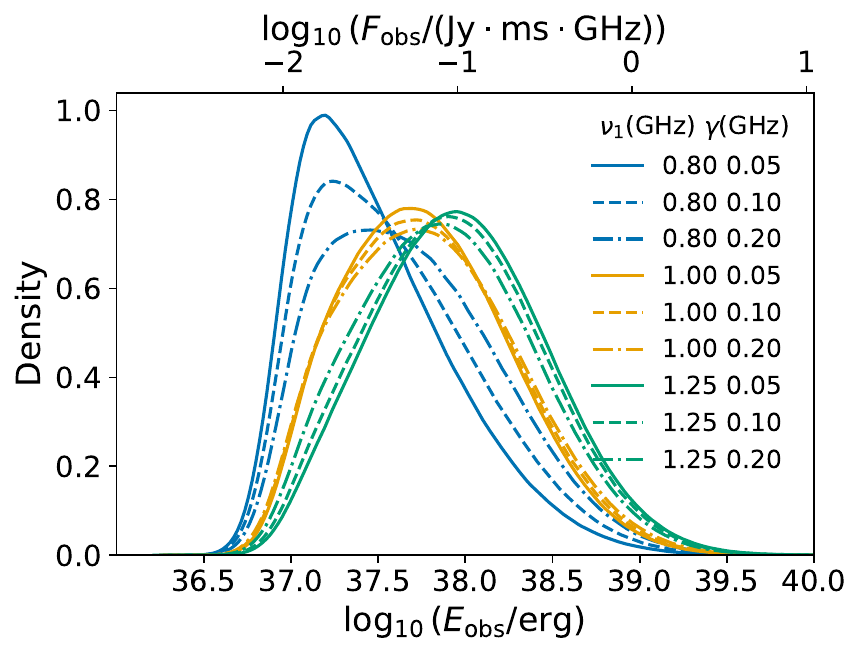}}
\subfigure[Exponential]{\includegraphics[width=2in]{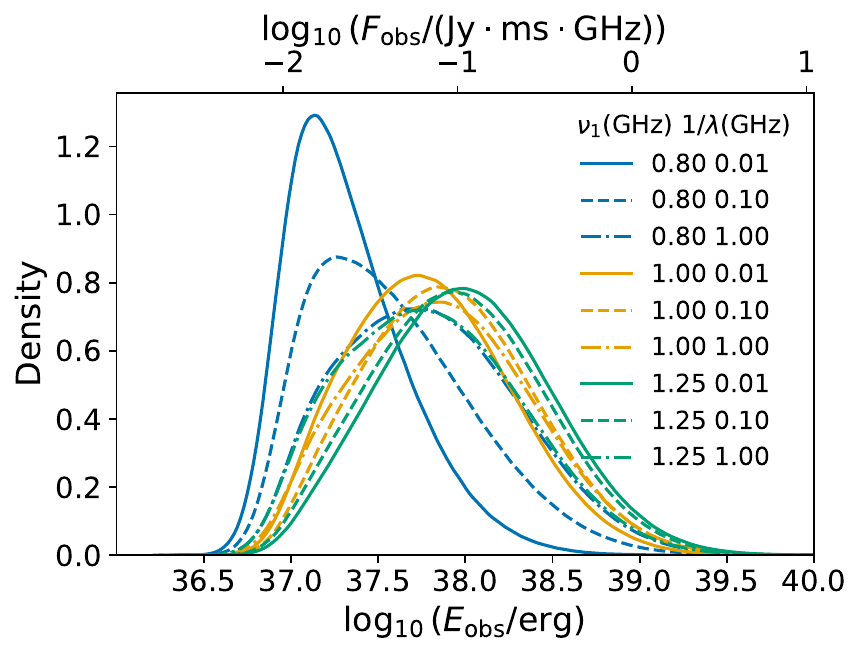}}
\caption{The PDFs of the observed energy distribution $p_{\mathrm{obs}}(E_{\mathrm{obs}})$ are depicted for different unimodal $\nu_{0}$ distributions. Each panel corresponds to a specific unimodal distribution, with the intrinsic energy function following a lognormal distribution.}
\label{fig: unimodal}
\end{figure*}

The results of the unimodal $\nu_{0}$ distribution cases, including the Gaussian distribution, the Lorentz distribution, and the exponential distribution, are presented in Figure \ref{fig: unimodal}.
These results are similar to the simulation of the constant $\nu_{0}$, where the observed energy function shifts towards lower energies as the center frequency deviates from the observing band edge, leading to a more pronounced impact of band selection effects. 
However, despite the presence of band selection effects, the resulting energy distribution is still unimodal, though its peak and shape change with $\nu_0$.
This indicates that converting an unimodal energy distribution into a bimodal one solely through selection effects alone is unlikely.
Similar results of the unimodal $\nu_{0}$ distribution were also found in previous works \citep{2022ApJ...941..127L,2024ApJ...966..115L}.

\subsection{Bimodal Central Frequency Distribution}
\begin{figure}
    \centering
    \includegraphics[width=0.96\linewidth]{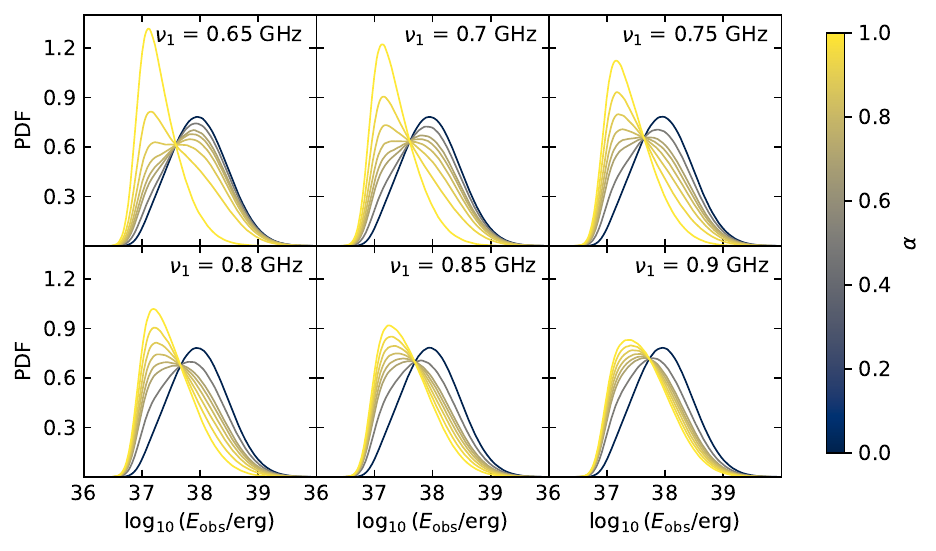}
    \caption{The PDFs of observed energy distributions $p_{\mathrm{obs}}(E_{\mathrm{obs}})$ in the case of two-Gaussian $\nu_{0}$ distribution ($\sigma_1 = \sigma_2 = 0.1$).
    Each subplot displays a series of $p_{\mathrm{obs}}(E_{\mathrm{obs}})$ for a fixed $\nu_1$.
    In each panel, these curves are plotted for different values of $\alpha$ ranging from 0.7 to 1 at intervals of 0.08.
    The special cases $\alpha = 0$, $0.5$, and $1$, which are also shown in the sub-figures.
    }
    \label{fig: wide}
\end{figure}

Next, we simulate the distribution of the observed energy $P_{\mathrm{obs}}(E_{\mathrm{obs}})$ using the two-Gaussian model.
The results are shown in Figure \ref{fig: wide}.
In this scenario, $\alpha = 0$ and $\alpha=1$ are two extreme cases, and the general $P_{\mathrm{obs}}(E_{\mathrm{obs}})$ is a linear combination of them.
The relative abundance of the two sub-samples determines the combined feature.
If one of the bases dominates the entire sample, it will also dominate the feature.
The band selection effects cause the observed energy distribution of the out-of-band sub-sample to exhibit a peak near the detection limit.

When $\nu_1$ is close to the lower edge of the observation band (e.g., $\nu_1 = 0.9 ~ \mathrm{GHz}$), the observed energy distributions corresponding to $\alpha = 0$ and $1$ exhibit negligible differences.
As a result, a distinct bimodal structure does not emerge.
In contrast, when $\nu_1$ lies farther away from the observing band, the energy distribution profiles for $\alpha = 0$ and $1$ differ significantly.
Thus, a bimodal pattern is likely to emerge when merging these two sub-samples.
Due to the limited observing band, the out-of-band sub-sample usually has a lower $E_{\mathrm{obs}}$ and more lost bursts, so the bimodal feature is restricted to the region where $\alpha > 0.5$.
Compared to the simulations of unimodal $\nu_{0}$ distribution, the key difference is that the bimodal $\nu_{0}$ distribution has more degrees of freedom to control the relative abundance of the out-of-band and in-band sub-samples, making the emergence of the bimodality easier. In conclusion, a bimodal pattern in $P_{\mathrm{obs}}(E_{\mathrm{obs}})$ can be generated by the band-selection effect, but this typically requires a non-trivial $\nu_{0}$ distribution.

\begin{figure}
    \centering
    \includegraphics[width=0.4\linewidth]{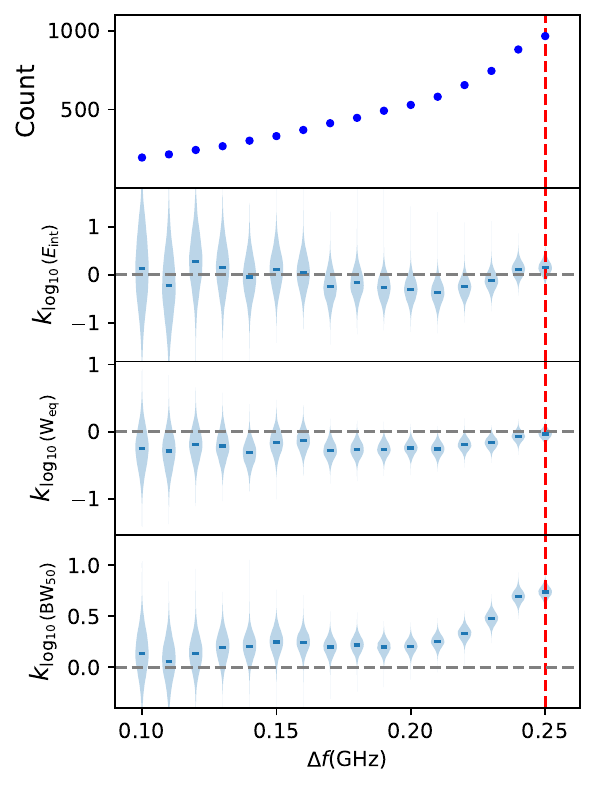}
    \caption{The linear fits of $k_{\log_{10}(E_{\mathrm{int}})}$, $k_{\log_{10}(\mathrm{W}_{\mathrm{eq}})}$ and $k_{\log_{10}(\mathrm{BW}_{50})}$ for bursts with different ranges of $\nu_0$. $\Delta f$ means that the linear fit is made using the bursts with $\nu_0$ ranging from $1.25 - \Delta f$ GHz to $1.25 + \Delta f$ GHz. The violin plots represent the posterior distribution and the blue bar is the mean value of the posterior. The red dotted line represents the edge of the observing band, corresponding to the band edge $\Delta f = 0.25$.}
    \label{fig: cor}
\end{figure}

\section{Population Features} \label{sec:pf}
The fitting results of the linear slope k of the intrinsic energy $E_{\mathrm{int}}$, the burst duration $\mathrm{W}_{\mathrm{eq}}$, and the burst bandwidth $\mathrm{BW}_{50}$ against the central frequency $\nu_0$ are shown in Figure \ref{fig: cor}.
In the observational data, these parameters generally follow a lognormal distribution; therefore, we adopted a logarithmic (base-10) transformation before further analysis.
When $\Delta f$ is small, the fitting errors are limited by the number of sub-samples and therefore consistent with a zero slope at the 1 $\sigma$ confidence level.
With the increase of $\Delta f$, the fitting errors gradually decrease, and the results are consistent with the previous results until reaching $\Delta f = 0.2$ GHz.
After $\Delta f$ exceeds 0.2 $\mathrm{GHz}$, these slopes undergo significant change, with $k_{\log_{10}(\mathrm{BW}_{50})}$ even in tension with the previous results.
This is likely due to the unreliable estimation of the morphological parameters.
Given that only partial spectra are observed, local features are likely to influence or even dominate the fitting process, causing them to be outliers of the population, which can also be seen in Fig. 6 of \cite{2023ApJ...955..142Z}.
Therefore, we argue that samples with a central frequency ranging from $1.05$ to $1.45$ $\mathrm{GHz}$ are minimally influenced by the poor fit and thus accurately reflect the true morphology, making them suitable for evaluating the population features and reconstructing the intrinsic energy distribution.

For this golden sample with $\nu_0$ ranging from $1.05$ to $1.45$ $\mathrm{GHz}$, we fit the $\log_{10}(E_{\mathrm{int}})$ vs. $\nu_0$ slope to be $-0.301 \pm 0.241$ at the 1$\sigma$ confidence level, which is slightly negative but still consistent with $k_{\log_{10}(E_{\mathrm{int}})} = 0$ at the 2$\sigma$ confidence level.
Therefore, we argue that $E_{\mathrm{int}}$ and $\nu_0$ of FRB 20220912A are nearly uncorrelated within the FAST observing band, which can rule out the possibility of $E_{\mathrm{int}}$-$\nu_0$ correlation-induced bimodality for this source.
However, both slopes $k_{\log_{10}(\mathrm{W}_{\mathrm{eq}})} = -0.238^{+0.093}_{-0.096}$ and $k_{\log_{10}(\mathrm{BW}_{50})} = 0.206^{+0.061}_{-0.065}$ respectively show slightly negative correlation and positive correlation above 2$\sigma$ confidence level, but still consistent with zero at the 3$\sigma$ confidence level, which indicates that there are bursts with slightly narrower bandwidth and longer duration in the higher frequency.
This tendency is also consistent with the recent observation of the Allen Telescope Array \citep{2024MNRAS.52710425S}.
It is worth noting that a similar trend also exists in other repeaters, such as FRB 20121102A \citep{2018ApJ...863....2G}, FRB 20221124A \citep{2022RAA....22l4001Z}, and FRB 20240114A \citep{2025arXiv251008367L}, indicating that this may be a universal feature across at least one subset of the population of repeaters. So far, there has been no explanation for this phenomenon.

\section{The Intrinsic Energy Distribution of FRB 20220912A} \label{sec:intE}
\begin{figure}
    \centering
    \subfigure{\includegraphics[width=3in]{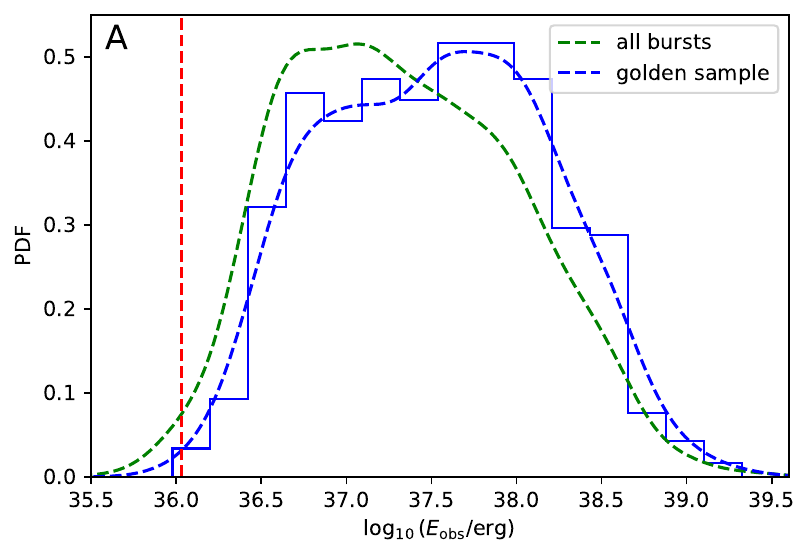}}
    \subfigure{\includegraphics[width=3in]{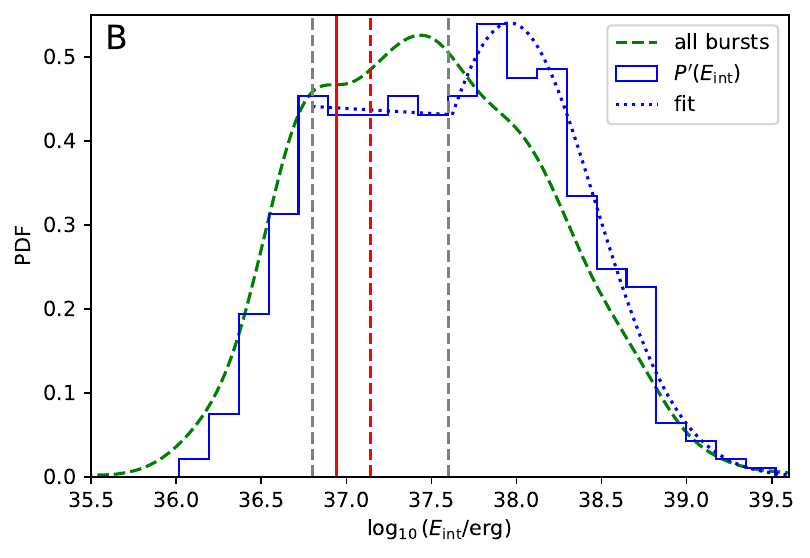}}
    \caption{The observed and intrinsic energy distributions of FRB 20220912A. \textbf{A}: The green dotted curve represents the kernel density estimation (KDE) of the observed energy of all 1067 bursts from \cite{2023ApJ...955..142Z}. The blue histogram and the KDE curve denote the observed energy distribution of our golden sample, and the red dotted line is the 90\% detection threshold $E_{90} = 1.07 \times 10^{36}$ erg. \textbf{B}: The green dotted curve represents the KDE of the intrinsic energy of all 1067 bursts, where the energy is corrected through their best-fit morphological parameters. The blue histogram and dotted curve represent the intrinsic energy distribution and its corresponding fit with a log-normal function and a power law function of our golden sample, respectively.
    The grey dashed lines represent the transitions between the two adjacent segments in $P'(E_{\mathrm{int}})$. The red dashed line is the 90\% detection completeness threshold $E_{90} = 1.38 \times 10^{37}$ erg, and the red solid line is the 90\% detection completeness threshold $E_{90} = 8.72 \times 10^{36}$ erg when we consider the correlation between morphological parameters in the simulation. Both simulations are carried out with $\nu_0=1.25$ GHz and $\mathrm{SNR}_{\mathrm{th}} = 7$.
    }
    \label{fig: Eint}
\end{figure}

The golden sample with $\Delta f = 0.2$ GHz is used to determine the distribution of $E_{\mathrm{obs}}$ and $E_{\mathrm{int}}$, as shown in Figure \ref{fig: Eint}.
For comparison, we also show the observed and intrinsic energy functions of the entire sample in this figure.
The completeness functions are nearly identical for $\nu_0$ within the observing band, and ill-fitting subsets have been removed; thus, merging bursts with different values of $\nu_0$ would not introduce any bias.
So the distribution of the intrinsic energy $p'(E_{\mathrm{int}})$ above the detection limit would follow the true energy function $p(E_{\mathrm{int}})$.
For $\Delta f = 0.2$ GHz, the sub-sample size can reach 546 (about 52\% of the total selected sample), and the distribution of $E_{\mathrm{int}}$ above the detection limit can not be fit with a single function, confirming its bimodal nature.
The low-energy tail is due to the SNR cut-off and therefore determined by the selection function as predicted by simulations.
For $E_{\mathrm{int}} > 3.16 \times 10^{37}$ erg, a log-normal function is used to fit the distribution, with a characteristic energy $\sim 8.13 \times 10^{37}$ erg.

We also use a power law function to fit the distribution ranging from $6.31 \times 10^{36}$ erg to $3.16 \times 10^{37}$ erg, and the power law slope is $-1.011 \pm 0.028$ at the 1$\sigma$ confidence level.
This suggests a uniform distribution of $\log_{10}(E_{\mathrm{int}})$, which is quite different from the prediction of the log-normal section. The simulation predicts a 90\% completeness threshold of $E_{90} = 1.38 \times 10^{37}$ erg under the assumption of independent morphological parameters, and this threshold reduces to $E_{90} = 8.72 \times 10^{36}$ when we use a three dimensinal lognormal function with covariance to model $\{ E_{\mathrm{int}}, \mathrm{W}_{\mathrm{eq}}, \mathrm{BW}_{50} \}$ with its free model parameters to match the dataset. Since SNR is proportional to the observed fluence $F_{\rm obs}$ and inversely proportional to the duration $\mathrm{W}_{\mathrm{eq}}$, the positive correlation between $E_{\mathrm{int}}$ and $\mathrm{W}_{\mathrm{eq}}$ makes lower energy bursts' SNR higher than the independent scenario. This further implies that an accurate determination of the 90\% completeness threshold $E_{90}$ relies critically on the features of the observed sample.
The application of some special search methods, such as deep learning-based search algorithms \citep{2025arXiv251007002W} and sub-band search techniques, can enhance completeness at lower energy levels, resulting in a threshold that is lower than prediction.

The distributions of the intrinsic energy $p'(E_{\mathrm{int}})$ exhibit significant deviations from the observed energy distribution reported by \cite{2023ApJ...955..142Z}.
The disappearance of the low-energy peak and the more dramatic decline at the high-energy tail stand out as key discrepancies.
These variations stem from our analysis correcting energy losses attributed to the limited observing band and excluding poorly fitting samples. The effect of band-limited energy loss is primarily reflected in the determinant of the Jacobian matrix $J(\boldsymbol{\theta})$. When the proportion of energy loss is significant, this correction becomes crucial. Our golden sample specifically excludes bursts whose central frequencies lie too far away from the central bandwidth. For such excluded bursts, the band-limited effect is pronounced, and their observed energies are generally much lower than those in the golden sample, making the illusion of more bursts in the low-energy part. In addition, the observational selection function $S(E_\mathrm{int},\boldsymbol{\theta})$ is of great importance. Correcting for this term would enable our intrinsic energy function to extend down to the lowest energies about $10^{36}$ erg. However, the current modelling of bursts in simulations remains relatively crude. Therefore, in this work, we have not accounted for this effect.

\section{Discussion} \label{sec:dis}
\subsection{Morphology}
When discussing the energy function, one has to bear in mind that the complexity of the burst morphology may hinder us from obtaining a robust estimation of the intrinsic morphological parameters, if only a partial segment of the burst falls within the observing band, thereby making the reconstruction of intrinsic energy unreliable. Part of the difficulty arises from the definition of the burst. 
There are three distinct definitions, namely the Combined Bursts Definition (CBD), the Separated Bursts Definition (SBD), and the cluster-burst.
The cluster-burst focuses on the burst morphology, which is defined as a group of individual bursts with a separation of less than 400 ms.
CBD takes a pulse profile above the detection threshold as a burst, potentially containing multiple adjacent sub-bursts defined by SBD.
Thus, the spectra of bursts under CBD may involve a combination of multiple Gaussian functions, leading to additional deviations when attempting to reconstruct the whole signal using a Gaussian model.

Another complication is that radio waves have multi-path effects.
The final signal at a particular time is the superposition of rays from paths with different plasma densities.
As the propagation speed is frequency dependent, the observed signal is left with an imprint of the medium in the astrophysical environments, and these phenomena are often referred to as scattering and scintillation.
Scattering typically alters the temporal characteristics, causing temporal tails, and scintillation generates frequency-dependent fluctuations.
Due to the combined effects mentioned above, reconstructing the entire signal with only a portion of the signal is extremely challenging. The small-scale fluctuations caused by thermal noise, scintillation, or residual RFI, seem to dominate the fitting process, leading to fitted shape parameters deviating from their true values.

\subsection{Comparison with Other Repeaters Observed by FAST}
FRB 121102 is the first repeater observed by FAST that exhibits a bimodal observed energy distribution \citep{2021Natur.598..267L}. Its bimodality has been further confirmed subsequently by long-term monitoring by FAST \citep{2025arXiv251007002W, 2025arXiv250318084H} and Arecibo \citep{2023MNRAS.519..666J}. The observed energy functions can usually be described by two log-normal functions or a two-segment power law, which is also appropriate for FRB 20201124A \citep{2022Natur.611E..12X, 2022RAA....22l4002Z}, and FRB 20220912A \citep{2023ApJ...955..142Z}. The broken power law model typically consists of two power law segments separated by a break point, where the high energy index is steeper than the low energy part. The intrinsic energy function of FRB 20220912A suggests the low energy power law index is $-1.011 \pm 0.028$, which is significantly different from $-1.38\pm0.02$ reported by \cite{2023ApJ...955..142Z}. Such a difference should be attributed to the choice of using observed energy or intrinsic energy to construct the energy function. When intrinsic energy is employed, the fraction of bursts in the low-energy region is reduced significantly.

In the FAST reports, the energy functions for several sources are typically constructed using the observed energy \citep{2021Natur.598..267L, 2022RAA....22l4001Z, 2022RAA....22l4002Z, 2022Natur.611E..12X, 2022Natur.611E..10N, 2023ApJ...955..142Z, Zhang:2025qzn}, this approach effectively circumvents errors introduced by morphological uncertainties. These energy functions often exhibit a peak at the low-energy end near the detection threshold. The consistency of this feature across different sources suggests that it is likely attributable to observational effects. Our work further validates this hypothesis: when bursts near the edge of the bandwidth are removed and the energy function is reconstructed based on intrinsic energy, this peak disappears. We highlight our work in understanding the feature in the energy function.

\subsection{Implications from Intrinsic Energy Functions}
Note that the detected distributions of $\nu_0$, $\mathrm{W}_{\mathrm{eq}}$, and $\mathrm{BW}_{50}$ could be biased, although we have assumed that there is no correlation between the morphological parameters in the simulation.
According to the simulation in Section \ref{sec:se}, the bias stems primarily from the bursts near the detection threshold.
In the power-law energy distribution used in \cite{2021ApJ...920L..18A}, lower-energy bursts are more abundant, leading to a higher fraction of missed bursts in the simulation.
Consequently, the observed distribution can be significantly affected in this case.
However, for energy distributions such as the log-normal or the intrinsic energy distribution of FRB 20220912A inferred in this work, the number of bursts in the low-energy regime is significantly reduced, resulting in only minor distortions in the observed distributions of morphological parameters.

We further simulate the band selection effect with different $\nu_0$ distributions, and demonstrate that it has the potential to generate a bimodal energy distribution.
However, even after mitigating the band selection effects in the FRB 20220912A data, the bimodal nature of the intrinsic energy distribution persists, suggesting that it is more likely to be associated with a fundamental mechanism, not merely the result of bias induced by the band selection effect.

The propagation effects, including some absorption processes and multi-path effects, can also modify the brightness of the bursts. Since the absorption processes typically are frequency-dependent, the observations of FRBs do not support such scenarios \citep{2023ApJ...956...67Y, 2024ApJ...974..160K, 2024A&A...685A..87W}.
Plasma lensing and scintillation, on the other hand, can narrow the spectrum and modulate the energy, which are more promising to produce a bimodal distribution than other propagation effects.
However, the event rates of these phenomena are quite low and therefore unlikely to account for the observed bimodal energy distribution \citep{2024ApJ...974..160K}. Theoretically, 
the gravitational self-lensing effect \citep{2024ApJ...973..123D} could amplify the brightness of some bursts, with an amplification probability that generally follows a power-law function. While this naturally explains the power-law decay trend in the high-energy regime of the energy function, it has difficulty accounting for other features in the energy functions, such as the dip at intermediate energies. Therefore, it is also considered an unlikely explanation.

The inconsistent predictions derived from the propagation effects and the observational data strongly suggest that the bimodal energy function is an intrinsic property related to the trigger and/or radiation mechanisms.
Currently, FRBs are thought to be generated within the magnetospheres of magnetars, which is supported by the observational evidence \citep{2020Natur.586..693L, 2024ApJ...972L..20N, 2024NSRev..12E.293J, 2025Natur.637...43M, 2025Natur.637...48N, 2025ApJ...988..175L}.
Starquake is the most promising trigger mechanism \citep{2018ApJ...852..140W, 2019MNRAS.488.5887S, 2019ApJ...879....4W, 2020RAA....20...56J, 2022MNRAS.517.4612L, 2023MNRAS.526.2795T, 2023ApJ...949L..33W, 2024MNRAS.530.1885T, 2024MNRAS.531L..57D, 2024RAA....24j5012W, 2025ApJ...988...62L}, where the released energy, potentially dissipated through faulting, sliding, or local density instabilities, may continuously contribute to driving subsequent, diverse types of bursts, and it is conceivable that there are different types of quakes which may produce the observed bimodality in the bursts. \cite{2025ApJ...979L..42W} proposed a starquake model to explain the features on the energy function in analogy with the earthquake's frequency-magnitude relation \citep{1956BuSSA..46..105G, 1992Natur.355...71P}. Such a model predicts that the low-energy power law index is -$(\beta+2)/3$. To match our fit, the power-law index $\beta$ for the fractured plate scale distribution is 1, which implies that the number of plates is inversely proportional to their size. This result should be interpreted with caution, since the connection between starquakes and FRBs currently remains theoretical. Further observational evidence is required to validate this hypothesis.

In addition to the trigger mechanism, the observed energy may also be affected by the beaming effect.
FRBs are commonly believed to be generated through coherent radiation by relativistic particles, with the relativistic effect causing only a fraction of the energy to be observed by observers on Earth.
This ratio is usually denoted by the beaming factor $f_b$. When considering the total energy budgets, the beaming factor of individual bursts cancels out and the global beaming factor is adopted \citep{2023RvMP...95c5005Z,2024RAA....24j5012W}.
A bimodal $f_b$ distribution also has the potential to transform a single energy distribution into a bimodal distribution.

The origin of the bimodal energy distribution remains an open question, and it is also possible that such a pattern may arise from the mixture of multiple radiation mechanisms \citep{2021Natur.598..267L}.

\section{Conclusions} \label{sec:con}
Statistical analysis provides us an effective way to study the engines and radiation mechanisms of FRB repeaters.
In this paper, we simulate the band selection effects due to the feature of the narrow spectra, and investigate two possibilities that the bimodality is induced by the $E_{\mathrm{int}}$-$\nu_0$ correlation and the band selection effects. We also analyze the population features and the intrinsic energy distribution in FRB 20220912A data from FAST observations.

When the central frequency is located within the observing band, the completeness is nearly 1 for burst energy above the detection limit, indicating that the detection is assured regardless of burst morphology. Conversely, if the central frequency is outside the observing band, the completeness decreases sharply even at high energy, leading to a peak in the observed energy distributions.
Simulations involving unimodal and bimodal central frequency distributions reveal that the band selection effects could contribute to the observed energy distribution, but in the sample of bursts observed by FAST, the band selection effect alone is insufficient to produce a bimodal energy distribution for an unimodal intrinsic energy distribution. However, in the case of the two-Gaussian $\nu_{0}$ model, the band selection effects can help generate the bimodal energy distributions seen in the data. 

From the analysis of FRB 20220912A, we find that in this case the energy and central frequency are independent, while the bandwidth shows a positive correlation, and the burst duration has a negative correlation with the central frequency.
From a statistical perspective, it indicates that bursts at different frequencies have similar energy but distinct kinds of morphology.
By removing inconsistent samples statistically with the sample near the center of the observing band, we reconstruct the intrinsic energy distribution.
The intrinsic energy distribution exhibits a log-normal function at high energies, with a component that differs from a log-normal distribution at lower energies. Compared with previous results, the low-energy peak vanishes, and the decline in the high-energy tail becomes steeper. These results confirm the bimodal nature of the energy function of this FRB repeater.

%% Also note that the akcnowlodgment environment does not support long amounts of text. If you have a lot of people and institutions to acknowledge, do not use this command. Instead, create a new \section{Acknowledgments}.
\begin{acknowledgments}
We thank Bing Zhang, Weiwei Zhu, Yongkun Zhang, and Yidan Wang for helpful discussions.
This work made use of data from the FAST FRB Key Science Project.
X.-H. L. acknowledges support from the National SKA Program of China (Nos. 2022SKA0110100 and 2022SKA0110101) and NSFC (grant No. 12361141814).
W.-Y. W. acknowledges support from the NSFC (No.12261141690 and No.12403058), the National SKA Program of China (No. 2020SKA0120100), and the Strategic Priority Research Program of the CAS (No. XDB0550300).
X.-L. C. acknowledges support from the NSFC (grant No. 12361141814).
\end{acknowledgments}

\bibliography{sample631}{}
\bibliographystyle{aasjournal}



\end{document}